\documentclass[preprint,proceedings]{rmaa}

\usepackage{paralist}

\usepackage{psfrag,color}

\SetYear{2004}
\SetConfTitle{COMPACT BINARIES IN THE GALAXY AND BEYOND}

\title{Bowen Fluorescence from Companion Stars in X-ray Binaries} 

\author{
  J. Casares,\altaffilmark{1} 
  D. Steeghs,\altaffilmark{2}
  R.I. Hynes,\altaffilmark{3}
  P.A. Charles,\altaffilmark{4} 
  R. Cornelisse,\altaffilmark{4} 
  and K. O'Brien\altaffilmark{5}}

\altaffiltext{1}{Instituto de Astrof\'\i{}sica de Canarias, La Laguna, Spain.}
\altaffiltext{2}{Harvard-Smithsonian Center for Astrophysics, Cambridge, USA.}
\altaffiltext{3}{The University of Texas at Austin, Austin, USA.}
\altaffiltext{4}{University of Southampton, Southampton, UK}
\altaffiltext{5}{European Southern Observatory, Santiago, Chile.} 

\shortauthor{J. Casares et al.}
\shorttitle{BOWEN FLUORESCENCE IN X-RAY BINARIES}

\fulladdresses{
\item J. Casares: Instituto de Astrof\'\i{}sica de Canarias, 38200 La Laguna, 
Tenerife, Spain 
  (\email{jcv@ll.iac.es}).
\item D. Steeghs: Harvard-Smithsonian Center for Astrophysics, 60 Garden 
Street, MS-67, Cambridge, MA 02138 
  (\email{dsteeghs@cfa.harvard.edu}).
\item R.I. Hynes: The University of Texas at Austin, Astronomy Department, 
1 University Station C1400, Austin, Texas 78712
  (\email{rih@obelix.as.utexas.edu}).
\item P.A. Charles: School of Physics \& Astronomy, University of Southampton, 
Southampton, SO17 1BJ, UK 
  (\email{pac@astro.soton.ac.uk}).
\item R. Cornelisse: School of Physics \& Astronomy, University of Southampton, 
Southampton, SO17 1BJ, UK 
  (\email{cornelis@astro.soton.ac.uk}).
\item K. O'Brien: European Southern Observatory, Casilla 19001, Santiago 19, 
Chile 
(\email{kobrien@eso.org}).
}

\listofauthors{J. Casares, D. Steeghs, R.I. Hynes, P.A. Charles, R. Cornelisse 
\& K. O'Brien}
\indexauthor{Casares, J.}
\indexauthor{Steeghs, D.}
\indexauthor{Hynes, R.I.}
\indexauthor{Charles, P.A.}
\indexauthor{Cornelisse, R.}
\indexauthor{O'Brien, K.}

\abstract{This paper will review a new technique of detecting
companion stars in LMXBs and X-ray transients in outburst using the
Bowen fluorescence lines at $\lambda\lambda$4634-4640. These lines are
very efficiently reprocessed in the atmospheres of the companion
stars, and thereby provide estimates of the $K_2$ velocities and mass
functions. The method has been applied to Sco X-1, X1822-371 and
GX339-4 which, in the latter case, provides the first dynamical
evidence for the presence of an accreting black hole. Preliminary
results from a VLT campaign on V801 Ara, V926 Sco and XTE J1814-338
are also presented.}

\addkeyword{accretion}
\addkeyword{accretion discs}
\addkeyword{X-rays: binaries}
\addkeyword{X-ray: stars}

\begin{document}
\maketitle

\section{Introduction}
\label{sec:intro}

The Galaxy is populated with $\sim$50 known {\bf persistent}, bright Low
Mass X-ray Binaries (LMXBs) whose optical emission is triggered by
X-ray reprocessing in the gas surrounding the compact object, mainly
the accretion disc. The companion star is $\sim~10^6$ times fainter
than the optical disc and hence completely undetectable. This has
hampered dynamical studies of LMXBs which have been restricted so far
to radial velocity studies of X-ray transients in {\bf quiescence}
(e.g. Charles \& Coe 2003). In several cases, the quiescent companion
spectrum is just too faint for current instrumentation (e.g. GX339-4,
N Oph 93)
or the target is contaminated by a bright line-of-sight star (e.g. Aql
X-1, 4U 2129+47). Dynamical studies and mass determination of compact
stars in LMXBs can yield new black hole discoveries and, more
importantly, probe for the existence of ``massive'' neutron stars. The
latter would rule out soft equations of state and give further support
that LMXBs are indeed the progenitors of millisecond pulsars, spun up
by accretion.

\begin{figure}[!t]
  \includegraphics[width=\columnwidth, height=5cm]{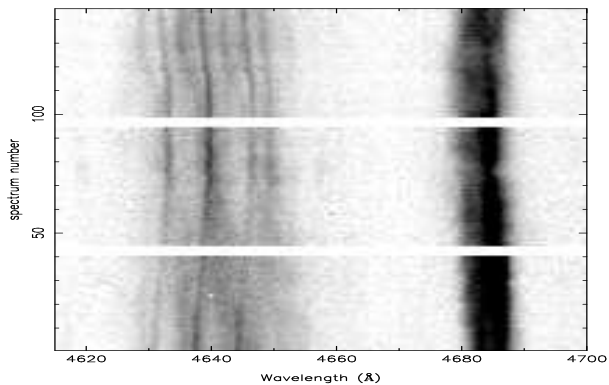}
  \caption{Trailed spectrogram of the Bowen blend and \ion{He}{II} 
  $\lambda$4686 line showing the Doppler shift of the narrow \ion{C}{III} 
  and \ion{N}{III} components. From Steeghs \& Casares (2002).}
  \label{fig:trail}
\end{figure}

\section{Detection of donor in Sco X-1}
\label{sec:scox1}

A new avenue for mass determination in LMXBs has been opened thanks to the 
discovery of narrow high-excitation emission components arising from the 
irradiated companion in Sco X-1 (Steeghs \& Casares 2002), the most prominent 
being \ion{C}{III} 4647-50 and \ion{N}{III} 4634-40 (see 
Fig.~\ref{fig:trail}). 

The \ion{N}{III} lines are produced by Bowen fluorescence through
cascade recombination (which requires \ion{He}{II} Ly$\alpha$ seed
photons), and must arise on the irradiated companion because they are
extremely narrow ($FWHM\leq50$ km s$^{-1}$), and move in anti-phase
with the compact star (as traced by the \ion{He}{II}
$\lambda$4686 wings).  Furthermore, the radial velocity curve is
sinusoidal indicating a fixed structure in the binary frame.  This
work represents the first detection of the companion star in Sco X-1
and opens a new window to extract dynamical information in a
population of $\simeq$ 20 LMXBs with optical counterparts.

\section{GX339-4 and X1822-371}
\label{sec:others}

The Bowen fluorescence diagnostic is a powerful technique also for transient 
sources as we have clearly demonstrated in Hynes et al. (2003). GX339-4 has 
been a black hole candidate for decades based on its X-ray properties but no
dynamical proof could be found. In summer 2002 we used the opportunity of a 
new outburst episode to obtain AAT, NTT and VLT spectroscopy which revealed 
(1) \ion{He}{II} velocities modulated with an orbital period of 1.76 d, and 
(2) narrow \ion{N}{III} Bowen components from the companion star with a 
velocity semi-amplitude of 317 $\pm$ 10 km s$^{-1}$. The implied mass function
is 5.8 $\pm$ 0.5 M$_{\odot}$ and is robust evidence for an accreting
black hole.  

\begin{figure}[!t]\centering
  \vspace{0pt}
  \includegraphics[angle=-90,width=0.94\columnwidth]{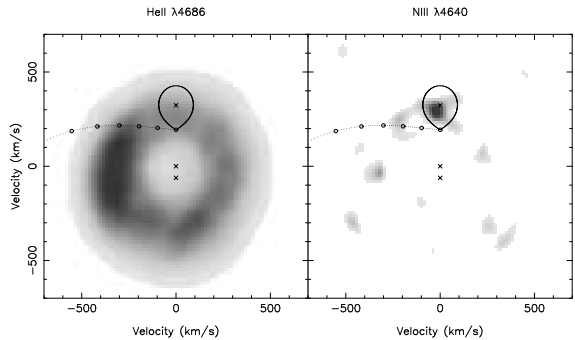}
  \caption{Doppler maps of \ion{He}{II} $\lambda$4686 and \ion{N}{III} 
$\lambda$4640 in 
  the pulsar LMXB X1822-371. The Roche lobe of the companion and gas stream
  trajectory are overplotted for a 1.4 M$_{\odot}$ neutron star. 
  From Casares et al. (2003).}
  \label{fig:doppler}
\end{figure}
  
The next obvious target is X1822-371 because, at B=16 it is one of the
brightest LMXBs. It is also a key system because it is eclipsing,
hence the inclination is well constrained, {\bf and} an X-ray pulsar, making
the orbit of the neutron star well constrained. In summer
2002 we obtained AAT spectroscopy but the moderate S/N prevented
identification of the narrow fluorescence components in individual
spectra.  However, we exploited the Doppler Tomography technique which 
simultaneously uses all the information contained in the phase-resolved 
emission profiles to reconstruct the emissivity distribution in velocity 
space (see Casares et al. 2003). Fig. 2 presents the Doppler
maps of \ion{He}{II} $\lambda$4686 and \ion{N}{III} $\lambda$4640. The
former displays a classic accretion disc distribution whereas the
latter shows a very compact spot consistent with the velocity and
phasing of the companion star.  The spot velocity, 300 km s$^{-1}$, is
a lower limit to the true velocity of the donor star because it is
formed on the inner irradiated face.  This, combined with the
knowledge of the inclination and neutron star's orbit, leads to solid
lower limits to the masses of the neutron star and companion of 1.14
$\pm$ 0.06 M$_{\odot}$ and 0.36 $\pm$ 0.02 M$_{\odot}$,
respectively. Tighter constraints require modelling the {\it
K-correction} (Wade \& Horne 1988) to determine the displacement of
the irradiated region from the center of mass of the donor star.
  
\section{VLT Survey}
\label{sec:vlt}

We have started a VLT survey of LMXBs to target new candidates with
strong Bowen emission for future studies. These are MM Ser, X1957+115,
LU TrA, V926 Sco, GX9+9, GR Mus, V801 Ara and X0614+091. In summer
2003 we observed V926 Sco, V801 Ara and a newly discovered transient,
the accreting millisecond pulsar XTE J1814-338 using VLT+FORS2. Our
Doppler tomograms enables us to detect the companion star in
\ion{N}{III} $\lambda$4640 and derive lower limits to their
K-velocities of 223, 282 and 345 km s$^{-1}$ respectively (Casares et
al. 2004a,b in preparation).

\begin{figure}[!t]
  \includegraphics[width=\columnwidth, height=6.8cm]{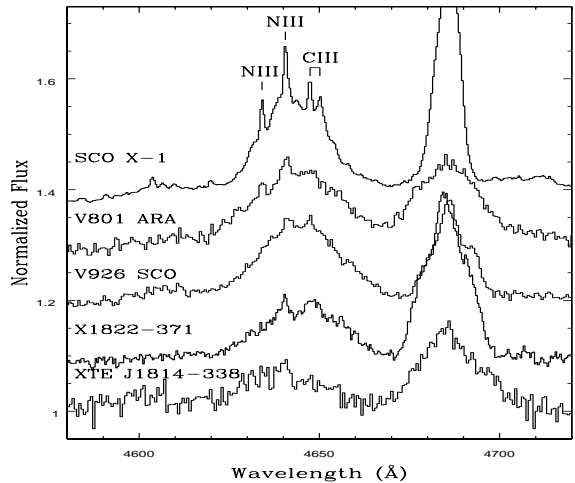}
  \caption{Doppler-corrected sums, in the rest frame of the companion, 
  of systems studied by our group so far.}
  \label{fig:bowen}
\end{figure}

%


\begin{thebibliography}

\bibitem{} Casares, J. et al. 2003, ApJ, 590, 1041
\bibitem{} Charles, P. \& Coe, M.J. 2003, astro-ph/0308020
\bibitem{} Hynes, R.I. et al. 2003, ApJ, 583, L95
\bibitem{} Steeghs, D. \& Casares, J. 2002, ApJ, 568, 273
\bibitem{} Wade, R.A.. \& Horne, K. 1988, ApJ, 324, 411

\end{thebibliography}
\end{document}